\documentstyle[aps,prl,twocolumn,epsf]{revtex}

\newcommand{\beq}{\begin{equation}}
\newcommand{\eeq}{\end{equation}}
\newcommand{\beqa}{\begin{eqnarray}}
\newcommand{\eeqa}{\end{eqnarray}}

\newcommand{\0}{\Delta_0}
\newcommand{\1}{\Delta_1}

\begin{document}

\draft
\twocolumn[\hsize\textwidth\columnwidth\hsize\csname @twocolumnfalse\endcsname
\title{Field induced $d_{x^2-y^2}+id_{xy}$ state and marginal stability of
high-Tc superconductors}

\author{  A.V. Balatsky}
\date{\today}

\address{T-Div and MST-Div, Los Alamos National Laboratory,
Los Alamos, NM 87545, USA}

%\hfill{DRAFT}

\maketitle
%\widetext
\begin{abstract}
%\leftskip 2cm
%\rightskip 2cm

It is shown that the   {\em complex} $d_{xy}$ component is
generated  in d-wave superconductor in  the magnetic field. As one enters
superconducting state at finite field  the
normal to superconducting transition occurs into  bulk  $d_{x^2-y^2}+i
d_{xy}$ state . The driving force for the
transition is the linear coupling between magnetic field and non zero
magnetization of the $d_{x^2-y^2}+i d_{xy}$ condensate. The external magnetic
field violates  parity and time reversal symmetries and the nodal quasiparticle
states respond by generating the $id_{xy}$  component of the order parameter,
with the magnitude
estimated to be on the order of few Kelvin.
Parity (P) and time reversal (T) symmetries  are violated in this state.

\end{abstract}
%\pacs{PACS numbers: 70.20.Pa, 73.20.Dx, 72.17.+a}
\noindent PACS numbers: 74.20.-z,   74.25.Dw
% \vskip2pc
\
\
]

Symmetry of the order parameter and existence of the gap nodes in high-T$_c$
superconductors has been one of the main experimental questions addressed for
the last few years. By now the majority of the data support the d-wave symmetry
of the order parameter \cite{wallman93,hardy93,shen93,Tsuei98}.
However,  recent experiments on  thermal transport in BiSrCuO superconductor
reveal a
number of  anomalies, prompting the suggestions that the secondary
superconducting order parameter is developed in the external magnetic field,
thus
lowering the symmetry of the initial $d_{x^2-y^2}$ to  order parameter that
contains more than one distinct components  \cite{Ong1,Aubin}.

Based on the data it was suggested that first order phase
transition from original $d_{x^2-y^2}$-wave  $(d)$  to $d_{x^2-y^2} +id_{xy}$
$(d+id')$ state occurs abruptly in the  magnetic field.  Theoretical approach,
describing this transition as an abrupt bulk generation of the secondary $id'$
component upon increased field was proposed \cite{Laugh1,Rama,Franz1}. It was also
found that the vortices  play an important role in the observed anomaly
\cite{Aubin,Kapit}.

Motivated by these experiments   I present here the alternative approach to the
field induced $d+id'$ transition in
high-$T_c$ superconductors. I focus on the high field region $H\simeq H_{c2}$
where vortices are so dense that they eventually destroy superconducting state.
In this regime I still find that   the d-wave state can be ``distorted'' by the
external field, producing $d+id'$ state in the bulk with the intrinsic orbital
moment, Figs.(1,2).

D-wave superconductor has low energy quasiparticles in the nodes of the d-wave
gap. These low lying states   have a vanishingly small gap
and hence can easily respond to the external perturbations \cite{s-wave}.  This
fact  opens up the
possibility to generate the second component of the order parameter, orthogonal
to
the initial d-wave state, from these ``normal'' quasiparticles at the nodes.
This softness of the d-wave state to the secondary component generation in the
presence of perturbations, such as  scattering on the surface
\cite{Green1,Sauls1} or the magnetic impurity scattering \cite{M,B1} is the
reflection of the   {\em marginal} stability of d-wave superconductors
\cite{BM}.

\begin{figure}
\epsfxsize=2in
\centerline{\epsfbox{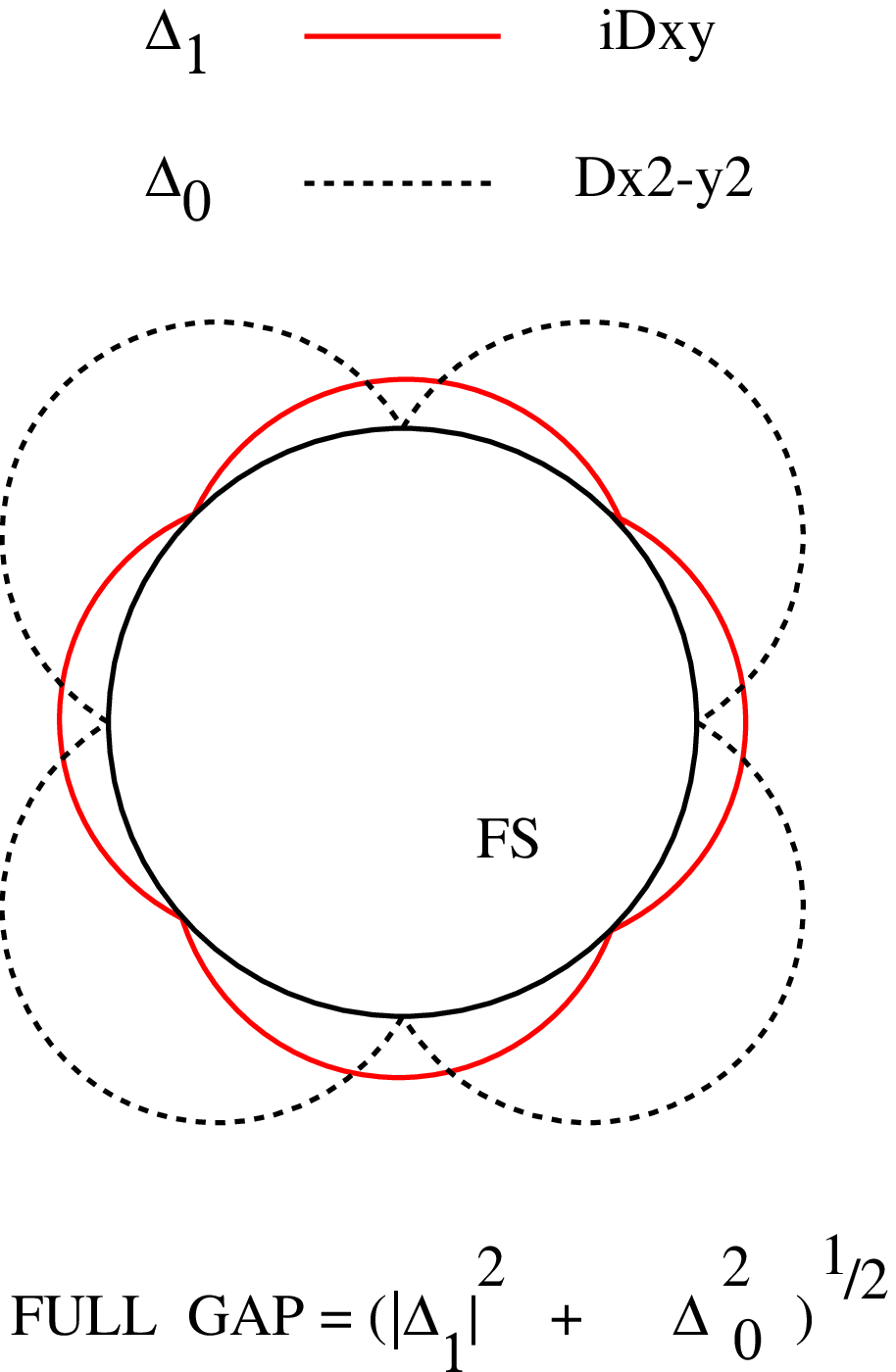}}
\caption[]{ Fermi surface (FS), taken to by cylindrical, is shown with the
angular dependent $\0(\Theta), \1(\Theta)$ gaps. The  $d+id'$ state is fully
gapped because for the complex order parameter the particle gap will depend on
sum of squares of real $(d)$ and imaginary $(d')$ amplitudes.}
\label{Fig1}
\end{figure}

Here I
consider the case of external magnetic field and argue that the normal to
superconducting transition in the field occurs into bulk $d+id'$ state: d-wave
state is {\em marginally stable} in the field.  Magnetic field $H||z$ violates P
and T symmetries.
The superconducting state responds to the field by generating the $id_{xy}$
component so that the total order parameter symmetry in the bulk becomes
$d+id'$.

The physical origin of the instability is the bulk magnetic  moment $\langle M_z \rangle$ in 
$d+id'$ state.   First I
present a qualitative argument for $id'$ induction by the
external field.   The relevant interaction is the
$\langle M_z \rangle B$
coupling   to the magnetic
field $B||z$, where $B$ is the induction in external field $H$.   In the pure
phase one can think of d-wave state as an equal
admixture of the orbital moment $L_z = \pm 2$ pairs:
\beqa
\Delta_0(\Theta) = \Delta_0 \cos 2\Theta ={\Delta_0\over{2}}(\exp(2i\Theta) +
\exp(-2i\Theta))
\eeqa
Here $\Theta$ is the 2D planar angle of the momentum on the Fermi surface,
$\Delta_0$ is the magnitude of the $d_{x^2-y^2}$ component. Motivated by the
layered   structure of the
cuprates, I consider 2D
$d_{x^2-y^2}$ superconductor. In the presence of the external field
$H$ the coefficients of the $L_z = \pm 2$ components will shift {\em linearly}
in $B$ with {\em opposite} signs:
\beqa
\Delta_0(\Theta) \rightarrow {\Delta_0\over{2}}((1 + \eta
B)\exp(2i\Theta)\nonumber\\
+ (1 - \eta B)\exp(-2i\Theta))
 = \Delta_0(\Theta) + i \ B \Delta_1(\Theta)
 \label{eta}
\eeqa
 where $\Delta_1(\Theta)  \propto  \eta \  \sin 2\Theta$ -- is the $d_{xy}$
component and $\eta$ is  the coupling constant. The  relative phase $\pi/2$ of
these two order
parameters comes out naturally  because  $d+id'$ state has a partially
noncompensated
orbital moment $L_z = +2$ . The pure
d-wave state has nodes of the gap  and generated $id_{xy}$ gap will ``seal'' the
nodes making  the state fully gapped, as is shown in Fig.1.

To show how complex $id_{xy}$ component appears, consider a
macroscopic Ginzburg-Landau (GL) functional which allows   the
linear coupling between the original $d_{x^2-y^2}$ order parameter
$\Delta_0(\Theta) = \Delta_0 \cos 2\Theta$ and the field
induced $d_{xy}$ component: $\Delta_1(\Theta) = \Delta_1 \sin
2\Theta$. The GL functional contains the {\em linear } coupling term:
\beqa
F_{int} = i{\eta\over{2}} \0 \1^* B + h.c.
\label{int2}
\eeqa
where $\eta   $ is the macroscopic coupling constant, Eq.(\ref{eta}).  This
coupling is possible only for $d+id'$ and not for $d+is$ symmetry of
the order parameter.   The question I     address here  is to find the
conditions when the second component is generated in the presence of the
external field.  I find that: a)  instability into $\0+i\1$ state occurs at
$T_c(H)$ greater than the transition temperature for the initial transition
$T^0_c(H)$ in the absence of the coupling $F_{int}$.   Upon
entering the superconducting state the symmetry of the condensate in the bulk is
$d +id'$ from the outset, as is indicated in Fig.2. This is the main result of
this paper. b)   $id_{xy}$ induced component is linear in the applied field
$\1/\0 \propto H$.  
\begin{figure}
\epsfxsize=3.5in
\centerline{\epsfbox{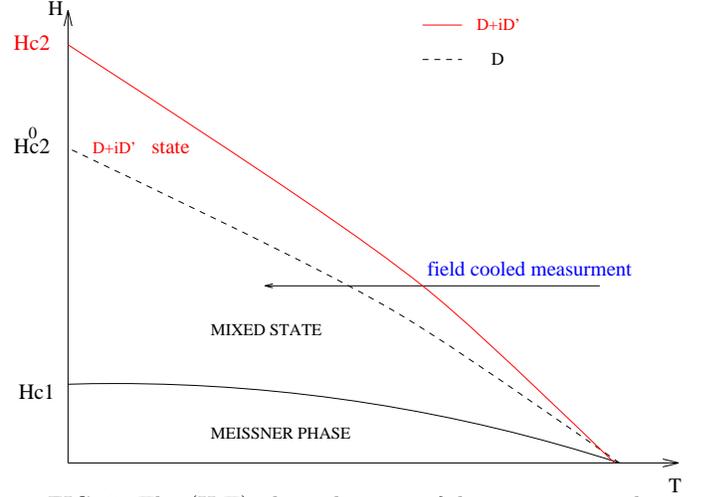}}
\caption[]{The (H,T) phase diagram of d-wave superconductor is shown. The dashed
line corresponds to the normal $\rightarrow$ superconductor transition in the
absence of the $\0\1$ coupling: $H^0_{c2}(T)$. The solid line is the new phase
boundary $H_{c2}(T)$ in the presence of the $\0\1^*$ coupling in the free
energy. The $d+id'$ state has higher transition temperature and hence the first
state to form below $H_{c2}$ line is the $d+id'$ state in the bulk.  The effect
of the induced $d+id'$ state can be seen
in
the field cooled experiment, as indicated. The slope of the $H_{c2}(T)$ is
linear at small $T$ since the shift $\delta T_c(H)$, Eq.(\ref{dTc}) is quadratic
in $H$. I ignore here the effects of possible vortex melted  phases. The new
$H_{c2}(T)$ is drawn out of scale  and is closer to the $H_{c2}^0(T)$ line. In
this theory  the $H=0$ transition occurs into   a pure d-wave state.}
\label{Fig2}
\end{figure}

I start with the GL free energy of the d-wave superconductor in the  vicinity
of $H_{c2}(T)$ field. The input fields are: $\0$ - $d_{x^2-y^2}$-wave order
parameter, $\1$ - $d_{xy}$ component  and external field along z- axis $H$. I
will consider the infinite  two dimensional case, where all the
gradients are in the $(xy)$ plane only. Here for simplicity  I  also consider
the case  of $d_{x^2-y^2} + id_{xy}$ admixture only, ignoring any other
components of the order parameter, that can be mixed in \cite{Mineev2}.
The reason is that  $d_{x^2-y^2} + id_{xy}$ is the only singlet state on the
square lattice   where both  P and  T are broken and
the condensate has nonzero  orbital momentum $\langle L_z\rangle \sim i \0\1^* +
h.c.$ \cite{Zeeman}.

The GL functional density  takes the form:
\beqa
F = \int dr [{\alpha_0\over{2}}(T - T^0_c)|\0(r)|^2 + {\beta_0\over{4}}|\0(r)|^4 +\nonumber\\K_{ij}|(D_i \0(r))(D_j\0(r)^*)|^2+{B^2\over{8\pi}}
+F_{int}+{\alpha_1 \over{2}}|\1(r)|^2 ]
\label{GL1}
\eeqa
where the  first three  terms describe the instability of the pure
$d_{x^2-y^2}$ - wave superconductor,  with $T^0_c$ being  the critical
temperature
in the absence of  the field. Tensor $K_{ij}$ is the gradient tensor, which I
do not need specify right now \cite{K}. I used  $D_j = \nabla_j -
i{e\over{c}}A_j, j = x,y$ in gradient energy, $A_i$- is the gauge potential.
The last term describes  the {\em positive} energy shift for the $d_{x^2-y^2}$
state in the presence of the induced secondary component $d_{xy}$.  The
gradient terms for the induced $d_{xy}$ component are ignored since I consider
the case
where $\1$ is small and the gradient terms will lead to the small
corrections\cite{grad}.
$F_{int}$  in Eq.(\ref{int2}) is discussed in more details  below. For general
discussion of the GL terms see \cite{LZ,M1}.

From GL functional Eq.(\ref{GL1}) it follows  that the coupling to the external
field, as far as the secondary  $id'$ component is considered, is linear in
$F_{int}$. Whereas the stiffness term (last term in Eq.(\ref{GL1})) is
quadratic. Therefore the linear term dominates at small $\1(r)$ and yields
nonzero equilibrium value for $\1(r)$.

One can easily see that the only second order gradient term that couples
$\Delta_0$ and $\Delta_1$ is :
\beqa
D_x\Delta_0 D_y\Delta^*_1 - D_y\Delta_0 D_x\Delta^*_1 + h.c.
\label{int1}
\eeqa
leading to, among other terms, $-i{e\over{c}} B
\Delta_0\Delta^*_1 + h.c.$, where commutator $[D_x,D_y] =
i{e\over{c}}B$. Since there is no other
terms that couple $\0 $ and $\1$ in the gradient terms, I can write the
coupling term without derivatives as is done in Eq.(\ref{int2}).

To show that   the terms in Eqs.(\ref{int1}, \ref{int2}) are indeed full
invariants, 
recall that the square lattice group $D_4$ has  irreps: $A_1$ (s-wave), $A_2
\sim z$ , $B_{1g} \sim x^2-y^2$ - $d_{x^2-y^2}$-wave representation, $B_{2g}
\sim xy$ - $d_{xy}$ representation, all of them one dimensional and one
two-dimensional representation: $E \sim (x,y)$ - $p$ wave \cite{LL}. Consider
for example $B_{1g}$ and $B_{2g}$ coupling in Eq.(\ref{int1}): the the product
of $\0\1^* \sim A_{2g}$. The direct product of two derivatives, transforming as
$E$ each, is a  sum of all four one dimensional representations:   $E\times E =
A_{1g} + A_{2g} + B_{1g} + B_{2g}$. The antisymmetric combination $D_xD_y -
D_yD_x = i{e\over{c}} H \sim A_{2g}$ and I find that true scalar can be formed
by taking the product of two $A_{2g}$ representations $\0 \1^*$ and
$[D_x,D_y]$. Hence the interaction term Eq.(\ref{int1}) is a scalar and allowed
in free energy. Eq(\ref{int2}) follows immediately. Alternatively, as indicated
above, one can see that the $\0+i\1$ state has a nonzero angular momentum state
with expectation value of the magnetic moment of the condensate $\langle
M_z\rangle \sim
i\0\1^* + h.c.$. Therefore there is a linear coupling between external field and
  magnetic moment $F_{int} \propto - \langle M_z \rangle B$, Eq.(\ref{int2}).

 Next,   I use the fact that for $H \simeq H_{c2}$ the local field $h(r)$ is homogeneous
and d-wave  order parameter takes the form $\0(r) = \0 f(r)$, where $f(r)$   is
a function of the position of the nodes of the order parameter, e.g. $f(r) =
\prod_i(z-z_i)exp(-|z_i|^2/4 \ell^2_H)$  where $z_i = x_i +iy_i$ are the nodes
of the order parameter for the case of pure d-wave with isotropic gradient term
$K_{ij} = K\delta_{ij}$ and $\ell_H = (c/2eH)^{1\over{2}}$ \cite{Tez}.  I will
not specify the form of $f(r)$ which should come out as a solution for
particular choice of the lattice of the nodes of the d-wave order parameter in
our case. Instead I  note that the  $\1$ field enters quadratically into GL
functional and  one can  integrate over $\1$ field, to  obtain the  effective
theory for $\0$.

 Treating $\1$ and $\1^*$ as independent variables the minimization of the Gibbs
energy $G = F -{BH\over{4\pi}}$,
$\partial_{\1}G = \partial_{\1^*}G = 0$ leads to:
\beqa
 \1(r) =  {\eta\over{i \alpha_1}}H \0(r) + O(|\0(r)|^2)
 \label{dxy}
 \eeqa
 hereafter I ignore  small difference  $B-H= -4\pi M_z \propto O(|\0(r)|^2)$ near $H_{c2}$, where $\0$ is small. After
substituting this result in  Eq.(\ref{GL1}) the
free energy for $\0$
acquires negative contribution $\delta F = - \int dr[\eta^2/2\alpha_1 H^2
|\0(r)|^2]$. I obtain the GL functional Eq.(\ref{GL1}) with renormalized
$T^0_c$, which is now field dependent:
 \beqa
 T^0_c(H)  = T^0_c + \delta T_c(H),  \ \delta T_c(H) =
{\eta^2\over{\alpha_0\alpha_1}} H^2
 \label{dTc}
 \eeqa
The standard next step is to solve the GL equations for the $\0(r)$, using the
ansatz $\0(r)  = \0 f(r)$ as discussed above \cite{LZ,M1,Tez}. One would find the
free
energy for the amplitude of the order parameter $\0$ (now $\0,\1$ are {\em
homogeneous fields} corresponding to the amplitudes of $d+id'$ order) with
the field dependent $T_c(H)$:
\beqa
F=\alpha(T,H)/2 |\0|^2+\beta(T,H)/4|\0|^4 , \1={\eta\over{i \alpha_1}}H\0
\label{GL3}
\eeqa
with $\alpha(T,H)=\alpha_0 (T- T_c(H) - \delta T_c(H))$.
The $\alpha(H,T) $ and $\beta(H,T)$ parameters will  be determined by the vortex
lattice structure. Here the vortex lattice in  not specified explicitly and I assume that
for given configuration one finds the solution.  I write the general
expression for thus determined $\alpha(H,T)\propto T - T_c(H)$ in the absence of
the $\0\1^*$ coupling. It follows from Eq.(\ref{GL3}) that this coupling will
shift amplitude instability to higher fields/temperatures $T_c(H) \rightarrow
T_c(H) + \delta T_c(H)$ as is shown in Fig.(\ref{Fig2}).

To estimate the amplitude of the induced $\1$  one can use the BCS theory for
the
coefficients in the GL energy Eq.(\ref{GL1},\ref{GL3}). These estimates  are
useful as  the order of magnitude estimates at best. The coefficient $\eta$ in
$F_{int}$ (in units of energy)  describes the $\langle M \rangle H$ interaction
with the orbital moment of the condensate and  I estimate $\eta = \mu_B N_0^2$
, where $\mu_B={e\hbar\over{2mc}}$ is the Bohr magneton,  $N_0 = 1/eV$ per unit
cell- is the density of states at the Fermi surface. $\alpha_1$ was shown to be
$\alpha_1\simeq N_0/2$ \cite{B1}. Using $H_{c2}
\simeq \phi_0/2\pi \xi^2$,  with $\xi = 20 \AA$ as the low temperature coherence
length of superconductor,  one gets $\eta \simeq N_0 (H/H_{c2}) (a/\xi)^2$,
where $a= 3.8 \AA$ is the atomic length in the Cu-O plane.  From these estimates
and Eq.(\ref{GL3}) it follows that:
\beqa
|{\1 \over{\0}}| \simeq  (H/H_{c2}) (a/\xi)^2 \simeq 10^{-2}
\label{dxyest}
\eeqa
This makes the amplitude of the induced component $|\1|$ to be on the order of
few $K^o$. From Eq.(\ref{dTc})   the shift of $T_c$ turns out to be much
smaller:
$\delta T_c/T^0_c \sim (H/H_{c2})(a/\xi)^4 \sim 10^{-3}$.

 Eqs.(\ref{dxy},\ref{dTc}) prove  the points that a) the transition into $d+id'$
state
indeed occurs at higher $T_c$ if the $\0\1^*$ coupling is present and b)
the field induced $\1/\0$ $xy$ component is linearly proportional to the field.
Whether the existence of the induced $id'$ component can be observed in
experiments, such as penetration depth $\lambda(H,T)$, is an interesting question
which is currently under investigation. It is possible that
the  secondary component will cause  changes in the  vortex lattice
symmetry at low temperatures.

 It is useful to note the  relation of  these results  to the previous work. Laughlin   argued for the  coupling term similar to $F_{int}$,
Eq(\ref{int2}), although the focus of  his work was   on the  different part
of the phase diagram $H_{c1} \ll H \ll H_{c2}$ \cite{Laugh1}. It was predicted
that the transition into $d+id'$ state to be of the {\em first} kind with the
induced component $\1
\propto H^{1\over{2}}$.  In another approach, Ramakrishnan recently suggested the alternative
mechanism to generate the
$d+id'$ state in the bulk \cite{Rama}. His approach is based on the
 finite $id'$ component generated near the vortex cores. As one increases the
vortex density the overlapped $id'$ patches eventually produce the bulk $id'$
order. The starting point there was the single vortex solution which is
inapplicable at   $H \simeq H_{c2}$. The relation between phases discussed  in
\cite{Laugh1,Rama} and present work is unclear at a moment. One possibility is
that  there is a crossover or phase transition that separates high field
$d+id'$ state,  discussed here, from possible $d+id'$ state at lower $H \ll
H_{c2}$ fields.

In conclusion, I argue that the normal-to-superconducting transition of the
d-wave superconductor occurs into $d+id'$ state with the  field induced
secondary component $id'$  with $\1/\0 \propto H$.  Even if the interactions
are
repulsive in   the $xy$ channel this field induced component is present at
finite fields $H\simeq H_{c2}$.  The driving force for the $id'$ induction is
the linear coupling between orbital moment of the $d+id'$ condensate with the
external magnetic field. Penetration depth $\lambda$ and vortex lattice
structure  could  be sensitive to the presence of $id'$ gap and might be used to
detect it.

I am grateful to D.H. Lee, V.P. Mineev and  M. Zhitomirsky for
the
useful discussions. This work was supported by the US DOE.

%\begin{thebibliography}{99}


\begin{references}
\bibitem{grad} In writing Eq.(\ref{dxy})  one can neglect gradient terms of the
type
$|D_x(y) \Delta_1|^2$.  Near $T_c$ they are proportional
to the ratio of the superconducting coherence length at $T=0$ $\xi$ to the GL
coherence length $\xi_{GL} \propto (T_c-T)^{-1/2}$: $(\xi_0/\xi_GL)^2 ~ (Tc-T)$,
 whereas the homogeneous
quadratic term has no such smallness $\alpha_1 \sim N_0$.
\end{references}
\end{document}